\documentclass[
  ,final            
  ]
  {aipproc}

\layoutstyle{6x9}

\newenvironment{cfigure}[1][tbp]{\begin{figure}[#1]\centering}{\end{figure}}

\newcommand{\pt}{p_{T}}

\newcommand{\et}{E_{_T}}

\newcommand{\gevcc}[1]{\ensuremath{#1~\mathrm{GeV}/c^{2}}}

\newcommand{\pb}{pb$^{-1}$}

\newcommand{\ttb}{t\bar{t}}

\newcommand{\ppb}{p\bar{p}}

\newcommand{\mtop}{M_{top}}

\newcommand{\wjj}{W\to jj}

\newcommand{\met}    {\mbox{$\protect \raisebox{.3ex}{$\not$}\et$}}

\newcommand{\measAStatJES}[3]{\ensuremath{#1~^{+#2}_{-#3}~\mathrm{(stat.+JES)}}}
\newcommand{\measAComb}[3]{\ensuremath{#1~^{+#2}_{-#3}~}}
\newcommand{\measAStatSyst}[4]{\ensuremath{#1~^{+#2}_{-#3}~\mathrm{(stat.)}\pm #4~\mathrm{(syst.\gevcc{)}}}}

\newcommand{\measStatSyst}[3]{\ensuremath{#1~\pm #2 ~\mathrm{(stat.)}\pm #3~\mathrm{(syst.\gevcc{)}}}}

\newcommand{\bjs}{$b$-jets}
\newcommand{\bj}{$b$-jet}

\newcommand{\bq}{\ensuremath{b}-quark}

\newcommand{\lumi}{$\int \mathcal{L} dt$}

\newcommand{\ljets}{\ensuremath{t\bar{t}\to l\nu q\bar{q}^\prime b\bar{b}}}
\newcommand{\dzero}{D\O}

\begin{document}

\title{Top Quark Mass and Properties at the Tevatron}

\classification{12.15.Ff; 12.15.Lk; 14.65.Ha}
\keywords      {Particle physics; hadron colliders; top quark; electroweak physics; heavy quark}

\author{Jean-Fran\c{c}ois Arguin\\ on behalf of the CDF and \dzero\ Collaborations}{
  address={Department of Physics, University of Toronto,
   60 St. George St., Toronto, M5S 1A7, ON, Canada}
}

\begin{abstract}
We present recent analyses of top quark properties performed at
Run~II of the Tevatron. Measurements of the top quark
mass, branching ratios and $W$ boson helicity inside top quark decays
are covered.
\end{abstract}

\maketitle


\section{Top Phenomenology at the Tevatron}

The top quark has been discovered only recently \cite{topdisco} due to its very 
large mass: $\mtop \approx \gevcc{175}$. Indeed, the top quark is easily the heaviest 
particle in the Standard Model (SM). 
This peculiar property brings several reasons
to study it. First, the Yukawa coupling of the top quark
is nearly one which could be a sign that it plays a special role in the origin
of mass beyond the Standard Model. Second, radiative corrections due
to top quark loops are often dominant in the prediction of precision observables
like the $W$ or Higgs boson mass. Finally, its large mass implies it has a very short 
lifetime ($\sim 10^{-25}$s), about an order of magnitude smaller than the hadronization 
time, which provides a unique opportunity to study a bare quark.

The top quarks are produced by the Tevatron, a $\ppb$ collider operating at a center-of-mass 
energy of 1.96 TeV for the current period of data-taking (Run II). They are produced 
predominantly in pairs ($\ttb$) via the strong interaction with a predicted cross-section 
of $6.7^{+0.7}_{-0.9}$ pb \cite{ttbar_xsec_theory}.
 Because the $V_{tb}$ CKM matrix element is nearly one under the 3 families SM assumption, 
the top quarks are predicted to decay $>$ 99.8\% of the time to a real $W$ boson and
a \bq. The two resulting $W$ bosons in turn decay either hadronically
or leptonically, defining the three channels of $\ttb$ events (with branching ratios in 
parenthesis): ``all-hadronic'' for two hadronic decays (46\%), ``lepton+jets'' for one leptonic and 
hadronic decays (29\%) and ``dilepton'' for two leptonic decays (5\%). By ``leptons'' we refer only
to electrons and muons, taus being generally too difficult to identify to be usually considered
in top quark properties measurements.

The $\ttb$ events reconstruction are performed by the CDF and D\O\ detectors that are
described in detail elsewhere \cite{detectors}. Typical event selections 
include identification of isolated leptons with high transverse momentum ($\pt$), 
large transverse missing energy ($\met$) due to the undetected neutrino(s) from $W$
boson decays, several high-$\pt$ jets and identification of \bjs\ ($b$-tagging) generally
using secondary vertex tagging. 
The integrated luminosity (\lumi) used for the measurements presented here vary from 
$\approx$~200--350~\pb. 

\section{Mass Measurements}

The current information on $\mtop$ comes from measurements performed
during Run~I of the Tevatron (with a luminosity of $\approx$ 100 \pb). The 
world average is 
$\mtop = \gevcc{178.0 \pm 4.3}$ \cite{runI_mtop_comb}. The final goal in Run~II is
to collect between 4-8 fb$^{-1}$ and reduce the uncertainty on $\mtop$ to $\approx$ \gevcc{2}.

A precise measurement of $\mtop$ is strongly motivated in the SM
because the radiative corrections to many precision observables
are  dominated by top quark loops. A famous example is the indirect 
constraint on the Higgs boson mass ($M_H$). The current constraint 
is $M_H = \gevcc{126^{+73}_{-48}}$ \cite{lepewwg}. More detailed discussions on 
the precision electroweak observables can be found in these proceedings (see 
contribution from P. Renton).

Before describing specific analyses, we first discuss general considerations.
One challenge of the $\mtop$ measurement
is to solve the combinatorics problem in the event reconstruction. For example
there are four jets in the final state in the lepton+jets channel (\ljets), 
resulting in 12 possible jet-parton assignments (since the two $W$ daughter jets
are interchangeable in the reconstruction). The problem is 
simplified by tagging \bjs\ as described above. Another limitation of the measurement 
is the large uncertainty in the modeling of the jet energy response, referred
to as the jet energy scale uncertainty (JES). This is generally the dominant 
systematic uncertainty, and is currently about \gevcc{3} for CDF 
and 5--\gevcc{6} for \dzero\footnote{The \dzero\ JES uncertainty is expected to go down near the 
CDF level in the next few months.}. 

CDF recently reported the most precise measurement of $\mtop$ to date. 
It has been performed in the lepton+jets channel using \lumi\ = 318 \pb\ of data. One novelty of this analysis
is the usage of the hadronic $W$ boson decays ($\wjj$) to improve the jet energy
scale uncertainty. Templates of reconstructed top quark and $W$ boson mass are created in MC events
as a function of the true top quark mass and the jet energy scale and compared with the data. 
The fit to the data yields $\mtop = \measAStatJES{173.5}{3.7}{3.6}\gevcc{}$ where both the 
statistical and JES uncertainties are included. The systematic uncertainties not including the JES 
are small (\gevcc{1.7}). The final 
result is $\mtop = \measAComb{173.5}{4.1}{4.0} \gevcc{}$. Figure~\ref{f:fitConsistencyMreco} shows the 
reconstructed top quark mass in the data with the best fits from the MC templates overlaid.
\begin{cfigure}
\resizebox{13.0cm}{7.9cm}{\includegraphics{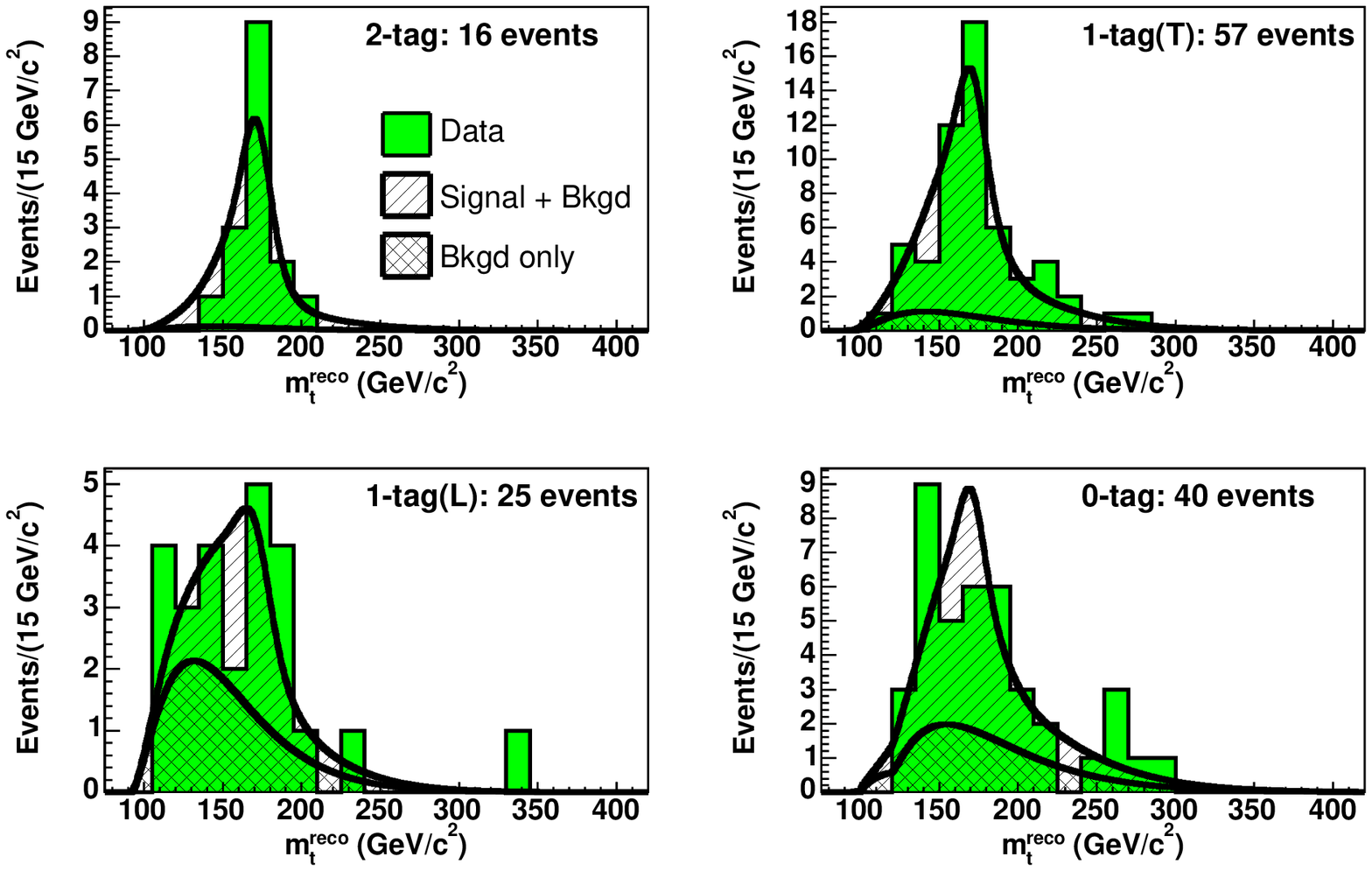}}
\caption{The reconstructed top quark mass for various subsamples of the lepton+jets sample
collected by CDF ($\sim$ 318 \pb).
Overlaid are the best fit from signal and background templates.}
\label{f:fitConsistencyMreco}
\end{cfigure} 
Analyses using the template method in the lepton+jets channel have also been performed by the 
\dzero\ collaboration. The best measurement 
so far corresponds to a luminosity
of 230 \pb\ and yields $\mtop = \measStatSyst{170.0}{4.2}{6.0}$.

A different class of analyses uses a matrix-element method that
consists of computing a probability for each event to be signal 
with a given top quark mass. The probability is computed using the full SM production
and decay matrix-elements.  This method has been shown to be very powerful statistically,
as demonstrated by the best measurement in Run I \cite{me_d0_runI}. 
The CDF collaboration has recently reported 
a measurement using this method: $\mtop = \measAStatSyst{173.8}{2.7}{2.6}{3.3}$. 
The result is in very good agreement with the template analysis above.
A similar measurement is expected to be released by the \dzero\ collaboration very soon.

The top quark mass can also be measured in the dilepton channel but with
larger statistical uncertainties due to the smaller branching ratio and the two neutrinos 
in the final state that complicate the event reconstruction.
The CDF collaboration has recently reported the best measurement in this channel to date 
using the matrix-element method:  $\mtop = \measStatSyst{165.3}{6.3}{3.6}$ (340 \pb). 
The best measurement at 
\dzero\ so far is:  $\mtop = \measAStatSyst{155}{14}{13}{7}$ (230 \pb).

\section{Other Properties}

It is very important to study in detail the properties of the top quark
since it might be more closely related to new physics than other SM particles 
due to its large mass. 
For instance, one can study the polarization of the $W$ boson 
inside the top decays. In the SM, 70\%, 30\% and 0\% of $W$ bosons are predicted
to have a longitudinal, left-handed and right-handed helicity, respectively. 
The $W$ helicity is sensitive to the angle between the charged lepton in $W$ rest frame 
and the \bj\ angle, denoted as $\cos\theta^*$. The \dzero\ collaboration
measured the fraction of right-handed $W$ bosons in top decays:
$f^+ < 0.25$ at 95\% confidence level (C.L.) \cite{whel_d0}. Figure \ref{f_whel_d0} shows
the distribution $\cos\theta^*$ in data. 
\begin{cfigure}
\resizebox{9.0cm}{5.5cm}{\includegraphics{cos_btag.epsi}}
\caption{Distribution of $\cos\theta^*$ for data
(points), signal assuming pure V-A interaction (full line), pure V+A interaction (dashed 
line) and background for events with $b$-tag (\dzero\ experiment \cite{whel_d0}).}  
\label{f_whel_d0}
\end{cfigure} 
The $W$ helicity is also sensitive to the charged lepton $\pt$ spectrum from $W$ boson decays
that has been used at CDF to measure the fraction of longitudinal $W$ bosons: 
$f^0 = 0.27^{+0.35}_{-0.21}$. 

The assumption that the top quark decays nearly 100\% of the time to $t\to Wb$ is 
checked by measuring $R = B(t\to Wb)/B(t\to Wq)$. This is done
by counting the number of $b$-tags in the sample. 
Such a measurement was performed at \dzero\ and yielded $R > 0.64$ at 95\% C.L..
A consistent result has been obtained by CDF: $R > 0.61$ at 95\% C.L.  \cite{br_cdf}.

Other properties of the top quark will be measured in Run II as
more data is accumulated like the electric charge, $\ttb$ spin correlation and rare decays.

\section{Conclusion}

The precise determination of top quark properties is already well underway in Run~II
at the Tevatron. The best measurement of the top quark mass to date has been performed at CDF:
$\mtop = \measAComb{173.5}{4.1}{4.0} \gevcc{}$. A very competitive measurement is expected
to be released by \dzero\ very soon. Measurements of other properties of top quarks
show a good agreement with the SM using only the small $\ttb$ samples collected so far
 ($\mathcal{O}(100)$ events per experiment).  We can expect these datasets to grow
by a factor of ten or so by the end of Run II, which will greatly improve
our knowledge of the peculiar particle that is the top quark.

We note that there was no time to present all analyses of top quark properties
performed at the Tevatron. More details  can 
be found on the public web pages of the CDF \cite{public_cdf} and \dzero\ 
\cite{public_d0} experiments.

\bibliographystyle{aipprocl} 





\end{document}